\documentclass[aps,prc,twocolumn,nofootinbib]{revtex4}
\usepackage{amsmath,graphicx}

\begin{document}
\title{Triangular flow in hydrodynamics and transport theory}
\author{Burak Han Alver}
\affiliation{
Laboratory for Nuclear Science, Massachusetts Institute of Technology,
Cambridge, MA 02139-4307, USA}
\author{Cl\'ement Gombeaud}
\author{Matthew Luzum}
\author{Jean-Yves Ollitrault}
\affiliation{
CNRS, URA2306, IPhT, Institut de physique theorique de Saclay, F-91191
Gif-sur-Yvette, France}
\date{\today}
\begin{abstract}
  In ultrarelativistic heavy-ion collisions, the Fourier decomposition
 of the relative azimuthal angle, $\Delta\phi$,
  distribution of particle pairs yields a large $\cos (3\Delta\phi)$ component,
  extending out to large rapidity separations $\Delta \eta>1$.  This
  component captures a significant portion of the ridge and shoulder
  structures in the $\Delta\phi$ distribution, which have been
  observed after contributions from elliptic flow are subtracted.  An
  average finite triangularity due to event-by-event fluctuations in the
  initial matter distribution, followed by collective flow,
  naturally produces a $\cos (3\Delta\phi)$ correlation.  Using ideal
  and viscous hydrodynamics, and transport theory, we study the
  physics of triangular ($v_3$) flow in comparison to elliptic
  ($v_2$), quadrangular ($v_4$) and pentagonal ($v_5$) flow.  We make
  quantitative predictions for $v_3$ at RHIC and LHC as a function of
  centrality and transverse momentum.  Our results for the centrality
  dependence of $v_3$ show a quantitative agreement with data
  extracted from previous correlation measurements by the STAR
  collaboration. This study supports previous results on the
  importance of triangular flow in the understanding of ridge and
  shoulder structures. Triangular flow is found to be a sensitive
  probe of initial geometry fluctuations and viscosity.
\end{abstract}


\maketitle

\section{Introduction}

Correlations between particles produced in ultrarelativistic heavy-ion
collisions have been thoroughly studied experimentally. Correlation
structures previously identified in proton-proton collisions have been
observed to be modified and patterns which are specific to
nucleus-nucleus collisions have been revealed.  The dominant feature
in two-particle correlations is elliptic flow, one of the early
observations at RHIC~\cite{Ackermann:2000tr}.  Elliptic flow leads to
a $\cos (2\Delta\phi)$ term in the distribution of particle pairs with
relative azimuthal angle $\Delta\phi$.  More recently, additional
structures have been identified in azimuthal correlations after
accounting for contributions from elliptic
flow.~\cite{Adare:2008cqb,Abelev:2008nda,Abelev:2008un,Alver:2008gk,Alver:2009id,Abelev:2009qa}.
An excess of correlated particles are observed in a narrow ``ridge''
near $\Delta\phi=0$ and the away side peak at $\Delta\phi=\pi$ is wider
in comparison to proton-proton collisions. For central collisions and
high transverse momentum triggers, the away side structure develops a
dip at $\Delta\phi=\pi$ with two ``shoulders'' appearing. These ridge and
shoulder structures persist for large values of the relative rapidity
$\Delta\eta$, which
means that they are produced at a very early
times~\cite{Dumitru:2008wn}.

It has been recently argued~\cite{Alver:2010gr} that both the ridge
and the shoulder are natural consequences of the triangular flow ($v_3$) 
produced by a triangular fluctuation of the initial distribution.
The purpose of this paper is to carry out a systematic study of
$v_3$ using relativistic viscous
hydrodynamics, which is the standard model for ultrarelativistic heavy-ion
collisions~\cite{Luzum:2008cw}.
We also perform transport calculations~\cite{Gombeaud:2007ub},
because they allow us to check the range of validity of viscous
hydrodynamics, and also because they provide
further insight into the physics.
Along with $v_3$, we also investigate $v_4$ (quadrangular flow) and
$v_5$ (pentagonal flow).
In Sec.~\ref{s:whyv3}, we recall why odd moments of the azimuthal
distributions, such as $v_3$, are relevant.
In Sec.~\ref{s:gaussian}, we study the general properties of
anisotropic flow induced by a harmonic deformation of the initial
density profile using hydrodynamics and kinetic theory.
In Sec.~\ref{s:predictionsv3}, we present
our predictions for $v_3$ and $v_5$ at RHIC and LHC.
The contribution of quadrangular fluctuations to $v_4$ is
difficult to evaluate because $v_4$ also has a large contribution from
elliptic flow~\cite{Borghini:2005kd}: this will be studied in a
forthcoming publication~\cite{Luzum:2011}.

\section{Correlations from fluctuations}
\label{s:whyv3}

A fluid at freeze-out emits particles whose azimuthal distribution
$f(\phi)$ depends on the distribution of the fluid
velocity~\cite{Borghini:2005kd}. $f(\phi)$ can generally be written as
a Fourier series
\begin{equation}
f(\phi)=\frac{1}{2\pi}\left(1+2\sum_{n=1}^{+\infty}v_n\cos(n\phi-n\psi_n)\right)
\label{fphi}
\end{equation}
where $v_n$ are the coefficients of anisotropic
flow~\cite{Voloshin:1994mz} which are real and positive, and $\psi_n$
is defined modulo $2\pi/n$ (for $v_n\not=0$).
Equivalently, one can write
\begin{equation}
\langle e^{in\phi}\rangle\equiv
\int_0^{2\pi}e^{in\phi}f(\phi)d\phi=v_n e^{in\psi_n},
\label{defvn}
\end{equation}
where angular brackets denote an average value over outgoing
particles.

Generally, $v_n$ is measured using the event-plane
method~\cite{Poskanzer:1998yz}.
However, two-particle correlation
measurements are also sensitive to anisotropic flow.
Consider a pair of particles with azimuthal angles $\phi_1$,
$\phi_2=\phi_1+\Delta\phi$. Assuming that the only correlation between the particles is due to the collective expansion,
Eq.~(\ref{defvn}) gives
\begin{equation}
\langle e^{in\Delta\phi}\rangle=\langle
e^{in\phi_1}e^{-in\phi_2}\rangle =\langle
e^{in\phi_1}\rangle\langle e^{-in\phi_2}\rangle =(v_n)^2.
\label{correlation}
\end{equation}
The left-hand side can be measured experimentally, and $v_n$ can thus
be extracted from Eq.~(\ref{correlation})~\cite{Wang:1991qh}.
Experimentally, one averages over several events. $v_n$ fluctuates
from one event to the other, and the observable measured through
Eq.~(\ref{correlation}) is the average value of $(v_n)^2$.
It can be shown that the event-plane method also measures the
RMS, $\sqrt{v_n^2}$, unless the ``reaction plane resolution''
is extremely good~\cite{Alver:2008zza, Ollitrault:2009ie}.

Most fluid calculations of  heavy-ion collisions are done
with smooth initial
profiles~\cite{Huovinen:2007xh,Hirano:2009ah,Song:2009rh,Broniowski:2009fm,Bozek:2009dw}. 
These profiles are symmetric with respect to the reaction plane
$\psi_R$, so that all $\psi_n$ in Eq.~(\ref{fphi}) are equal to $\psi_R$
(with this convention, all $v_n$ are not necessarily positive).
For symmetric collisions at midrapidity, smooth profiles are also
symmetric under $\phi\to\phi+\pi$, so that all
odd harmonics $v_1$, $v_3$, etc. are identically zero.
However, it has been shown that fluctuations in the positions of
nucleons within the colliding nuclei may lead to significant
deviations from the smooth profiles event-by-event~\cite{Miller:2003kd,Manly:2005zy}.
They result in lumpy
initial conditions which have no particular symmetry, and this
lumpiness should be taken into account in fluid dynamical
calculations~\cite{Gyulassy:1996br,Socolowski:2004hw,Holopainen:2010gz,Werner:2010aa}.
More precisely, one should calculate the azimuthal distribution for
each initial condition, then average over initial conditions.

Initial geometry fluctuations are a priori important for all $v_n$, as anticipated
in Ref.~\cite{Mishra:2007tw}.
Their effect on flow measurements has already been considered for elliptic flow
$v_2$~\cite{Andrade:2006yh,Alver:2006wh} and quadrangular flow
$v_4$~\cite{Gombeaud:2009ye}.
Event-by-event elliptic flow fluctuations have been measured and found
to be significantly large, consistent with the fluctuations in the
nucleon positions~\cite{Alver:2010rt}.
Directed flow, $v_1$, is constrained by transverse momentum
conservation which implies $\sum p_t v_1(p_t)=0$ and will not be
considered here.
In this paper, we study triangular flow $v_3$~\cite{Alver:2010gr}, and
pentagonal flow $v_5$, which arise solely due to initial geometry fluctuations.

\section{Flow from harmonic deformations}
\label{s:gaussian}

\begin{figure}[ht]
\includegraphics[width=0.4\linewidth]{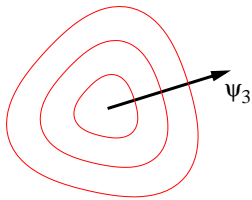}
\caption{(Color online) Contour plots of the energy density
  (\ref{gaussian}) for $n=3$ and $\varepsilon_3=0.2$.}
\label{fig:contour}
\end{figure}

Elliptic flow is the response of the system to an initial distribution
with an elliptic shape in the transverse plane
$(x,y)$~\cite{Ollitrault:1992bk}.
In this article, we study the response to higher-order deformations.
For sake of simplicity, we assume in this section that the
initial energy profile in the transverse plane $(x,y)$
is a deformed Gaussian at $t=t_0$:
\begin{equation}
\epsilon(x,y)=\epsilon_0\exp\left(-\frac{r^2\left(1+\varepsilon_n\cos(n(\phi-\psi_n)\right)}{2\rho^2}\right),
\label{gaussian}
\end{equation}
where we have introduced polar coordinates $x=r\cos\phi$, $y=r\sin\phi$. In
Eq.~(\ref{gaussian}), $n$ is a positive integer, $\varepsilon_n$
is the magnitude of the deformation, $\psi_n$ is a reference
angle, and $\rho$ is the transverse size.
Convergence at infinity implies $0\le\varepsilon_n<1$.
Fig.~\ref{fig:contour} displays contour plots of the energy density for
$n=3$ and $\varepsilon_3=0.2$. The sign in front of $\varepsilon_n$ in
Eq.~(\ref{gaussian}) has been chosen such that $\psi_n$ is the
direction of the flat side of the polygon. For $n=2$, it is the
minor axis of the ellipse, which is the standard definition of the
participant plane~\cite{Manly:2005zy}

For $t>t_0$, we assume that the system evolves according to the
equations of hydrodynamics or to the Boltzmann transport equation, until
particles are emitted, and we compute the azimuthal distribution
$f(\phi)$ of outgoing particles. 
The initial profile (\ref{gaussian}) is symmetric under the
transformation $\phi\to\phi+\frac{2\pi}{n}$, therefore $f(\phi)$ has the same
symmetry.
The only nonvanishing Fourier coefficients are $\langle
e^{in\phi}\rangle$, $\langle e^{2in\phi}\rangle$, $\langle
e^{3in\phi}\rangle$, etc. 
Symmetry of the initial profile under the transformation
$(\phi-\psi_n)\to -(\phi-\psi_n)$ implies
\begin{equation}
\langle e^{in\phi}\rangle=v_n e^{in\psi_n},
\label{defvn2}
\end{equation}
where $v_n$ is real. As we shall see below, $v_n$ is usually positive
for $\varepsilon_n>0$, which means that anisotropic flow develops
along the flat side of the polygon (see Fig.~\ref{fig:contour})

We now present quantitative results for $v_n$, as defined by
Eq.~(\ref{defvn2}), using two models.
The first model is relativistic hydrodynamics (see~\cite{Luzum:2008cw}
for details). We fix $\epsilon_0$, $t_0$ and the freeze-out
temperature to the same values as for a
central Au-Au collision at RHIC with Glauber initial
conditions~\cite{Luzum:2008cw}, and $\rho=3$~fm,
corresponding roughly to the rms values of $x$ and $y$.
Unless otherwise stated, results
are shown for pions at freeze-out. Corrections due to resonance
decays~\cite{Sollfrank:1991xm} are not included in this section. They
are included only in our final predictions in Sec.~\ref{s:predictionsv3}.
The second model is a relativistic Boltzmann equation for massless
particles in 2+1 dimensions (see~\cite{Gombeaud:2007ub} for details).
The only parameter in this calculation is the Knudsen number
$K=\lambda/R$, where the mean free path $\lambda$ and the transverse
size $R$ are defined as in~\cite{Gombeaud:2007ub}.
$\rho$ in Eq.~(\ref{gaussian}) is the rms width of the energy distribution,
while $R$ is defined from the rms widths $\sigma_x$ and
$\sigma_y$ of the {\it particle\/} distribution
by $R^{-2}=\sigma_x^{-2}+\sigma_y^{-2}$.
For a two dimensional ideal gas of massless particles,
the particle density $n$ is related to the energy density through
 $n\propto \epsilon^{2/3}$, which gives
 $R=\frac{\sqrt{3}}{2}\rho$.
Boltzmann transport theory is less realistic than hydrodynamics for several
reasons:
\begin{itemize}
\item{the equation of state is that of an ideal gas, while the
equation of state used in hydrodynamics is taken from lattice QCD:
it is much softer around the transition to the quark-gluon plasma.
Although transport is equivalent to ideal hydrodynamics when the mean
free path goes to zero, our results from transport and ideal
hydrodynamics differ in this limit, because of the different equation
of state.} 
\item{there is no longitudinal expansion.}
\item{particles are massless.}
\end{itemize}
The main advantage of transport theory is that
it can be used for arbitrary values of the mean free path, while
hydrodynamics can only be used if the mean free path is small. Furthermore,
the time evolution of the system can be studied
and no modeling is required for the freeze-out process using transport approach,
since one follows all elastic collisions until the
very last one.

\subsection{$v_n$ versus $\varepsilon_n$}

\begin{figure}[ht]
\includegraphics[width=\linewidth]{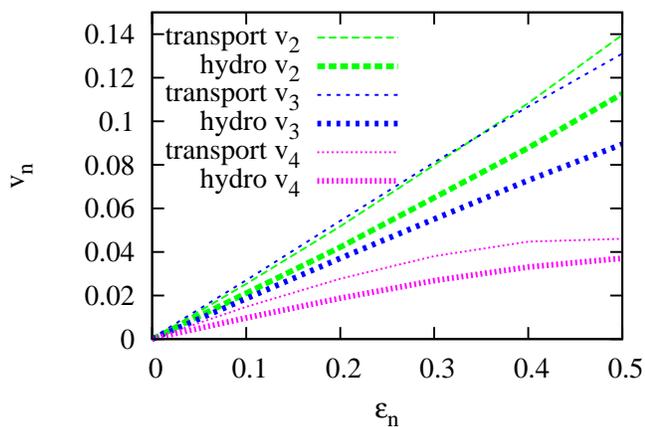}
\caption{(Color online) $v_n$ versus $\varepsilon_n$ in transport
theory and ideal hydrodynamics. The Knudsen number
in the transport calculation is $K=0.025$, close to the ideal
hydrodynamics limit $K=0$.
}
\label{fig:vnvsepsilon}
\end{figure}
Fig.~\ref{fig:vnvsepsilon} displays $v_n$ versus $\varepsilon_n$ for
$n=2,3,4$ in transport theory and ideal hydrodynamics (zero viscosity).
The values of
$v_n$ are smaller in hydrodynamics, which is due to the
softer equation of state~\cite{Bhalerao:2005mm}.

As expected from previous studies of $v_2$~\cite{Sorge:1998mk} and
$v_3$~\cite{Alver:2010gr}, we observe that $v_n$ is linear for small values of $\varepsilon_n$.
Non linearities are stronger for larger values of $n$, both in
transport theory and hydrodynamics.
A possible interpretation of these strong nonlinearities is that the
the contour plot of the initial density is no longer convex if
$
\varepsilon_n> 2/(n^2-2).
$
The threshold values for $n=3,4$ are $\varepsilon_3=\frac{2}{7}$ and
$\varepsilon_4=\frac{1}{7}$.
If the contour plot is not convex, the
streamlines (which are orthogonal to equal density contours) are no
longer  divergent: shock waves may appear, which hinder the
development of anisotropies.

The results presented in the remainder of this section are obtained in
the linear regime where $v_n\propto\varepsilon_n$.
In this regime, we find $v_2/\varepsilon_2\simeq 0.21$, in agreement
with other calculations~\cite{otherhydro}.
Note that in our hydrodynamic calculation, chemical equilibrium is
maintained until freeze-out. When chemical freeze-out is implemented
earlier than kinetic freeze-out, $v_2/\varepsilon_2$ is slightly
larger~\cite{Huovinen:2007xh}. 
Fig.~\ref{fig:vnvsepsilon} shows that $v_3/\varepsilon_3$ has a
magnitude comparable to $v_2/\varepsilon_2$, while $v_4/\varepsilon_4$
is significantly smaller. Our results for $v_5/\varepsilon_5$ (not
shown) are even smaller.

\subsection{Time dependence}

\begin{figure}[ht]
\includegraphics[width=\linewidth]{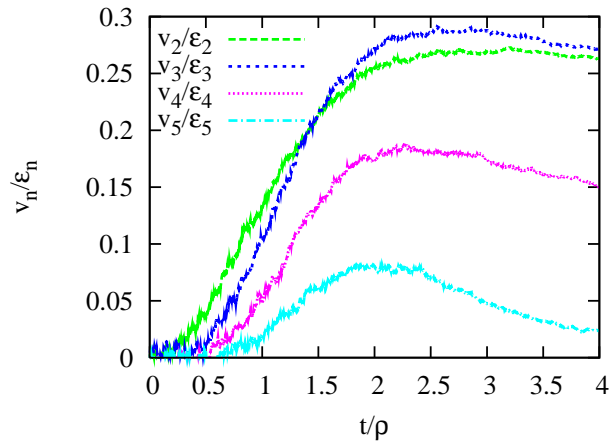}
\caption{(Color online) $v_n/\varepsilon_n$ versus time in transport
  theory. Each curve is the result of a single Monte-Carlo simulation
  with $K=0.025$ and $\varepsilon_n=0.1$. The number of particles in
  the simulation is $N=4\times 10^6$, and the corresponding
  statistical error on $v_n/\varepsilon_n$ is $3.5\times 10^{-3}$. }
\label{fig:vnvst}
\end{figure}
In the transport approach, one follows all the trajectories of the
particles, so that $v_n$ is well defined at all times, which is not
the case in hydrodynamics before freeze-out.
Fig.~\ref{fig:vnvst} displays the results for $v_n$ versus $t/\rho$,
where $\rho$ is the width of the initial distribution,
Eq.~(\ref{gaussian}).
As expected for dimensional reasons~\cite{Bhalerao:2005mm}, anisotropic
flow appears for $t\sim\rho$. However, $v_n$ appears slightly later
for larger $n$. This can be traced to the behavior of $v_n$ at early
times.
The transport results presented in Fig.~\ref{fig:vnvst} are obtained with
a very small value of the Knudsen number, $K=0.025$, close to the
ideal hydrodynamics limit.
In ideal fluid dynamics, the fluid transverse velocity
increases linearly with $t$, and $v_n$ involves a $n^{\rm th}$ power
of the fluid velocity, so that $v_n$ scales like $t^n$.
In transport theory, the number of collisions
increases like $t$ at early times, which gives an extra power of $t$,
and $v_n$ increases like $t^{n+1}$~\cite{Gombeaud:2007ub}.
In both cases, the behavior of $v_n$ at small $t$ is flatter for
larger values of $n$, which is clearly seen in Fig.~\ref{fig:vnvst}.

While elliptic flow keeps increasing with time (it slightly decreases
at later times, not shown in the figure), $v_n$ with $n\ge 3$
reaches a maximum and then decreases. The decrease is more
pronounced for larger $n$:
The mechanism producing $v_n$ is self quenching.

\subsection{Differential flow}

\begin{figure}[ht]
\includegraphics[width=\linewidth]{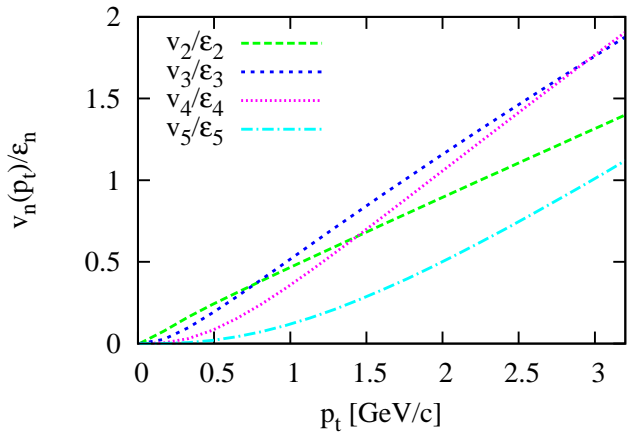}
\caption{(Color online) $v_n/\varepsilon_n$ versus $p_t$ in ideal
  hydrodynamics, with $\varepsilon_n=0.1$.
}
\label{fig:vnvspt}
\end{figure}

Fig.~\ref{fig:vnvspt} displays the differential anisotropic flow
$v_n(p_t)$ versus the transverse momentum $p_t$ for pions in ideal
hydrodynamics, scaled by the initial eccentricity $\varepsilon_n$.
At low $p_t$, one generally expects $v_n$ to scale like $(p_t)^n$ for
massive particles~\cite{Dinh:1999mn}\footnote{There is no such
constraint for massless particles where the $p_t\to 0$ limit is
singular. Our transport calculations for massless partons give
$v_n(p_t)\propto p_t$ at low $p_t$ for all $n$.}.
One clearly sees that $v_n$ is much flatter at low $p_t$ for larger
values of $n$.
For larger values of $p_t$, $v_n(p_t)$ is linear in $p_t$.
The arguments that explain this linear
dependence for $v_2$~\cite{Borghini:2005kd} can be generalized to
arbitrary $n$~\cite{Mishra:2008dm}.
The linear behavior at larger $p_t$ is also clearly seen in
Fig.~\ref{fig:vnvspt}. It has already been noted for
$v_3$~\cite{Alver:2010gr}.

The value of $v_3$ increases with $p_t$, which explains why the ridge
and shoulder are more pronounced with a high $p_t$ trigger (``hard''
ridge)~\cite{Putschke:2007mi}. Though the relative strength of $v_3$,
is smaller at low $p_t$, it is still comparable to $v_2$, leading to
the smaller``soft'' ridge~\cite{Daugherity:2008su}.
Predictions for $v_3(p_t)$ in viscous hydrodynamics for
identified particles are presented in Sec.~\ref{s:predictionsv3}.

\subsection{Viscous damping of $v_n$}
\label{s:viscous}

\begin{figure}[ht]
\includegraphics[width=\linewidth]{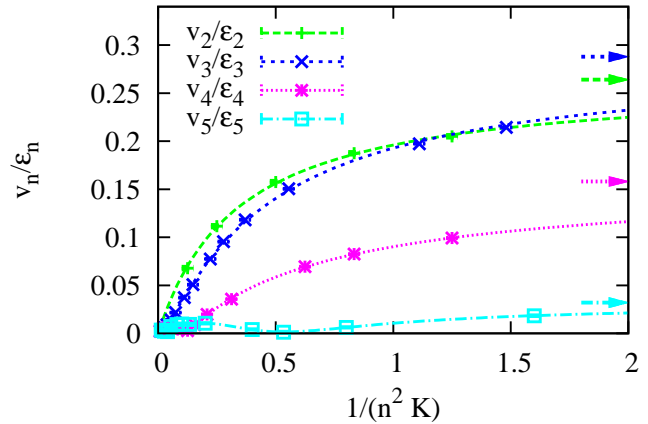}
\caption{(Color online) $v_n/\varepsilon_n$ versus $1/(n^2K)$ in transport
  theory. Values of $\varepsilon_n$ are  $\varepsilon_2=\frac{5}{13}$,
  $\varepsilon_3=\varepsilon_4=0.3$ and
  $\varepsilon_5=0.1$. Arrows indicate our extrapolation to $K=0$ (ideal
  hydrodynamics limit) using Eq.~(\ref{pade}). }
\label{fig:voloshin}
\end{figure}
We study the effect of viscosity first in the transport approach, then
in viscous hydrodynamics.
In transport, the degree of thermalization is characterized by the
Knudsen number $K$.
Experimentally, $1/K$ scales like $(1/S)(dN/dy)$, where $dN/dy$
is the multiplicity per unit rapidity, and $S$ is the overlap area
between the colliding nuclei~\cite{Drescher:2007cd}.
The dependence of $v_n$ on $K$ can be
studied by varying the collision system and the centrality of the collision~\cite{Voloshin:1999gs}

Transport is equivalent to ideal hydrodynamics in the limit $K\to 0$.
For small $K$, observables (such as $v_n$, or particle spectra)
deviate from the $K=0$ limit by corrections which are linear in
$K$. These are the viscous corrections: both $K$ and the shear
viscosity $\eta$ are proportional to the particle mean free path
$\lambda$.
Viscous damping is expected to scale with the wave number $k$
like $k^2$. Here, the wavelength of the deformation
is $2\pi R/n$, hence $k\sim n/R$.
Therefore viscous corrections should scale with $K$ and $n$
approximately like $n^2K$~\cite{Gavin:2010}.
The limit $K\to\infty$ (free streaming) is also interesting, since
$v_n$ vanishes in this limit. For large $K$, 
one therefore expects $v_n$ to scale like $1/K$, which is essentially
the number of collisions per particle~\cite{Gombeaud:2007ub}.
For intermediate values of $K$ ($K\sim 1$), no universal behavior is
expected, and observables depend on the scattering cross section used
in the transport calculation (dependence on energy and scattering angle).

Fig.~\ref{fig:voloshin} displays the variation of $v_n/\varepsilon_n$
versus the scaling variable $1/(n^2K)$ in the transport calculation.
Our numerical results can be fitted by smooth rational functions (Pad\'e
approximants)~\cite{Nagle:2009ip} for all $K$:
\begin{equation}
v_n(K)=v_n^{\rm ih}
\frac{1+B_nK+D_nK^2}{1+(A_n+B_n)K+C_nK^2+E_nK^3},
\label{pade}
\end{equation}
where $v_n^{\rm ih}$, $A_n$, $B_n$, $C_n$, $D_n$ and $E_n$ are fit
parameters. This formula has the expected behavior in both $K\to 0$
and $K\to\infty$ limits.
For $n=2$, the lowest-order formula, with
$B_2=C_2=D_2=E_2=0$, gives a good fit~\cite{Gombeaud:2007ub}.
For $n=3$, we obtain a good fit with using the next-to-leading order
approximant, with $D_3=E_3=0$ but free $B_3$, $C_3$.
For $n=4$ or $5$, we need all 6 parameters to achieve a good fit.
Fits are represented as solid lines in
Fig.~\ref{fig:voloshin}, and extrapolations to $K=0$ are indicated by
arrows.
As already noted above,
the hydrodynamics limits $v_3^{\rm ih}/\varepsilon_3$ and
$v_2^{\rm ih}/\varepsilon_2$ are comparable, while
$v_4^{\rm ih}/\varepsilon_4$ is smaller by roughly a factor of 2.
$v_5^{\rm ih}/\varepsilon_5$ is found to be further smaller by about a factor 5,
with a large theoretical uncertainty.

For small $K$, $v_n(K)\simeq v_n^{\rm ih}(1-A_n K)$:
the parameter $A_n$ measures the magnitude of the viscous correction.
Our fit gives $A_2=1.4\pm 0.1$~\cite{Gombeaud:2007ub}, $A_3=4.2\pm
0.3$, $A_4=11.0\pm 0.9$. The error bar on $A_5$ is too large to
extract a meaningful value.
For $n=2,3,4$, we observe $A_n\propto n^\alpha$ with $\alpha=2.8\pm
0.2$, closer to $n^3$ than to the expected $n^2$.
The fact that viscous corrections are larger for larger $n$ also
implies that the range of validity of viscous hydrodynamics is
smaller for $v_n$ with $n\ge 3$ than for $v_2$. Even after
rescaling $K$ by $n^2$, corrections are linear in
$K$ only for very small $K$, which is why higher-order
Pad\'e approximants are needed.

\begin{figure}[ht]
\includegraphics[width=\linewidth]{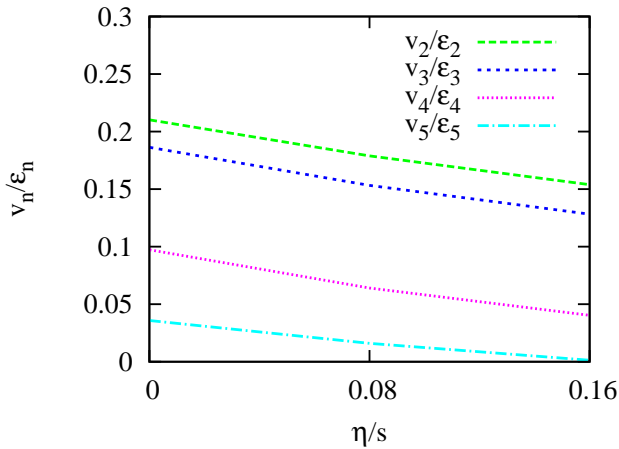}
\caption{(Color online) $v_n/\varepsilon_n$ versus $\eta/s$ in
hydrodynamics. The initial and freeze-out temperature are $T_i = 340$~MeV and $T_{f}=140$~MeV, respectively.}
\label{fig:eta}
\end{figure}
The magnitude of viscous effects can be seen more directly by varying
the shear viscosity $\eta$ in viscous
hydrodynamics~\cite{Heinz:2009cv}. For each value of 
$n$, we have performed three calculations with $\eta\simeq 0$ (ideal
hydrodynamics), $\eta/s=0.08\simeq 1/4\pi$~\cite{Kovtun:2004de},
and $\eta/s=0.16$, where $s$ is the entropy density.
The result is presented in Fig.~\ref{fig:eta}.
The variation of $v_n$ with $\eta$ is found to be linear for all $n$ 
for this range of viscosities, which is a 
hint that viscous hydrodynamics (which addresses
first-order deviations to local equilibrium) is a reasonable
description.
Interestingly, the lines are almost parallel, which means
that the absolute viscous correction to $v_n/\varepsilon_n$ depends
little on $n$.
However, since $v_n/\varepsilon_n$ is smaller for larger $n$, the
{\it relative\/} viscous correction is larger for larger $n$.
From the transport calculation, we expect that the relative viscous
correction is 3 times larger for $v_3$ than for $v_2$, and 8 times
larger for $v_4$ than for $v_2$. The increase in Fig~\ref{fig:eta} is
more modest.
Note that we keep the freeze-out temperature constant for all values
of $\eta/s$. Strictly speaking, this is inconsistent. Freeze-out is
defined as the point where viscous corrections become so large that
hydrodynamics breaks down: when the viscosity goes to zero, so does
the freeze-out temperature~\cite{Borghini:2005kd}.
By varying only $\eta/s$ and keeping $T_f$ constant, we only capture
part of the viscous correction~\footnote{We have checked that
  $v_3/\varepsilon_3$ is larger with a
  lower freeze-out temperature $T_f=100$~MeV. In particular, we find
  $v_3/\varepsilon_3>v_2/\varepsilon_2$, in agreement with the
  transport calculation.}.
Since triangular flow, like elliptic flow, develops at early times,
$v_3$ is sensitive to the value of $\eta/s$ at the high-density phase
of the collision.

\section{Predictions for $v_3$ at RHIC and LHC}
\label{s:predictionsv3}

\subsection{Triangularity fluctuations}

We now give realistic predictions for $v_3$ at RHIC and LHC.
The transport calculations in Ref.~\cite{Alver:2010gr} show that even
with lumpy initial conditions, $v_3$ in a given event scales
like the triangularity $\varepsilon_3$.
We define $\varepsilon_n$ as in~\cite{Alver:2010gr}:
\begin{equation}
\varepsilon_n e^{in\psi_n}\equiv -\frac{\int \epsilon(x,y)r^2
    e^{in\phi}dxdy}{\int \epsilon(x,y)r^2dxdy},
\label{defepsilon3}
\end{equation}
where $\epsilon(x,y)$ is the initial energy density and $(r,\phi)$ are
the usual polar coordinates, $x=r\cos\phi$, $y=r\sin\phi$.

Following the discussion in Sec.~\ref{s:whyv3}, experiments measure
the average value of $(v_n)^2$, so that
\begin{equation}
v_n^{\rm exp}=\sqrt{\langle (v_n)^2\rangle}.
\label{v3exp}
\end{equation}
Assuming $v_n=\kappa\varepsilon_n$ in each event, the measured $v_n$
scales like the root mean square $\varepsilon_n$ defined by
\begin{equation}
\varepsilon_n^{\rm rms}\equiv \sqrt{\langle (\varepsilon_n)^2\rangle}
\label{defrms}
\end{equation}
We compute $\varepsilon_n^{\rm rms}$ using two different models. The
first model is the PHOBOS Monte-Carlo Glauber
model~\cite{Alver:2008aq}, where it is assumed that the initial energy
is distributed in the transverse plane in the same way as nucleons
within colliding nuclei. We modify the initial model
slightly~\cite{Gombeaud:2009ye} 
by giving each nucleon a weight $w=1-x+xN_{\rm coll}$, where $N_{\rm
  coll}$ is the number of binary collisions of the nucleon. We take
$x=0.145$ at RHIC and $x=0.18$ at LHC~\cite{Bozek:2010wt}.
The second model is the Monte-Carlo KLN model of Drescher and
Nara~\cite{Drescher:2007ax}, which is the only model incorporating
both saturation physics and fluctuations.
Both of these models yield event-by-event eccentricity fluctuations,
which are consistent with measured elliptic flow fluctuations~\cite{Alver:2010rt}.
We loosely refer to the two models as Glauber and Color Glass Condensate
(CGC).

\begin{figure}[ht]
\includegraphics[width=\linewidth]{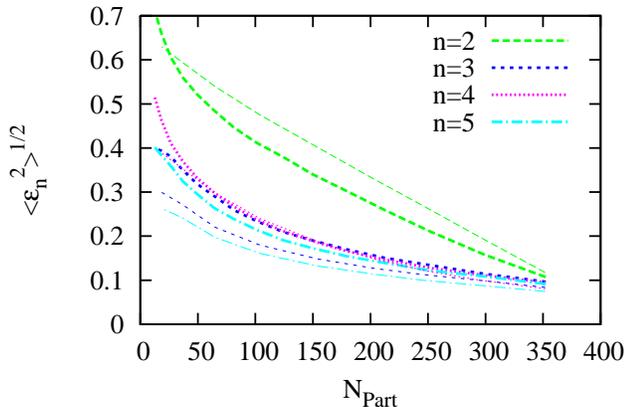}
\caption{(Color online)
Root mean square eccentricities $\varepsilon_n^{\rm
  rms}$ for $n=2,3,4,5$ for Au-Au collisions at 200 GeV per nucleon,
versus the number of participant nucleons $N_{\rm
  Part}$. $N_{\rm  Part}$ is used as a measure of
the centrality in nucleus-nucleus collisions: it is largest for central
collisions, with zero impact parameter~\cite{Miller:2007ri}.
Thick lines: Monte-Carlo Glauber model~\cite{Alver:2008aq}; Thin
lines: Monte-Carlo KLN model~\cite{Drescher:2007ax}.
}
\label{fig:epsilon}
\end{figure}
Fig.~\ref{fig:epsilon} displays
$\varepsilon_n^{\rm rms}$ as a function of the number of participants.
$\varepsilon_2^{\rm rms}$ is larger than $\varepsilon_{3,4,5}^{\rm rms}$ for
non-central collisions, which is due to the almond shape of the
overlap area.
The eccentricity is somewhat larger with CGC than
Glauber~\cite{Lappi:2006xc}.
$\varepsilon_3^{\rm rms}$ is very close to
$\varepsilon_5^{\rm rms}$.
Both vary with $N_{\rm Part}$ essentially like $(N_{\rm
  Part})^{-1/2}$, as generally expected for statistical
fluctuations~\cite{Bhalerao:2006tp}.
Unlike $\varepsilon_2^{\rm rms}$, they are slightly {\it smaller\/}
for CGC than for Glauber.
Since the only source of fluctuations that is considered in both
models is the position of the nucleons in the colliding nuclei,
this difference may be due to the technical implementation
of the Monte-Carlo KLN model.
Finally, $\varepsilon_4^{\rm rms}$ is slightly larger than odd
harmonics for peripheral collisions because the almond shape induces a
nonzero $\varepsilon_4$ as a second order effect.
Fig.~\ref{fig:epsilon} only displays results for Au-Au collisions at
RHIC. Results for Pb-Pb collisions at the LHC are similar, except for
a different range in $N_{\rm part}$, and a somewhat larger difference
between Glauber and CGC  for $\varepsilon_3$.

\subsection{Method for obtaining $v_3$ in hydrodynamics}

In order to make predictions for $v_3$, we start from a smooth initial
energy profile $\epsilon(r,\phi)$, possessing the usual symmetries
$\phi\to -\phi$ and $\phi\to\phi+\pi$. We then put by hand a $\cos
(3\phi)$ deformation through the transformation, inspired by Eq.~(\ref{gaussian}),
\begin{equation}
\epsilon(r,\phi)\to\epsilon\!\left(r\sqrt{1+\varepsilon'_3\cos(3(\phi-\psi'_3))},\phi\right),
\label{deformation}
\end{equation}
where $\varepsilon'_3$ is the magnitude of the deformation, and
$\psi'_3$ the flat axis of the triangle.
We choose $\varepsilon'_3=\varepsilon_3^{\rm rms}$.
The choice of $\psi'_3$ is arbitrary.
The initial profile has a nonzero eccentricity for
noncentral collisions, due to the almond shape of the overlap area.
Through Eq.~(\ref{deformation}), we add a triangular deformation to an
ellipse.
Since the original profile has $\phi\to\phi+\pi$ symmetry, $\psi'_3$
is equivalent to $\psi'_3+\frac{\pi}{3}$. Furthermore, $\psi'_3$ is equivalent
to $-\psi'_3$ due to $\phi\to-\phi$ symmetry. Therefore, one need only
vary $\psi'_3$ between $0$ and $\frac{\pi}{6}$. We choose the values
$0$, $\frac{\pi}{12}$ and $\frac{\pi}{6}$.

We then compute $\varepsilon_3$ and $\psi_3$ defined by
Eq.~(\ref{defepsilon3}).
With the gaussian profile (\ref{gaussian}), the input and output
values are identical: $\varepsilon'_3=\varepsilon_3$,
$\psi'_3=\psi_3$.
Our predictions use two sets of profiles which both describe RHIC data
well~\cite{Luzum:2008cw}: optical Glauber
and (fKLN) CGC. With both profiles,
$\varepsilon'_3$ differs from $\varepsilon_3$ by a few percent.
$\psi_3$ is essentially identical to $\psi'_3$, which means that the
elliptic deformation does not interfere with the triangular
deformation.
According to the previous discussion, we should tune $\varepsilon'_3$
in such a way that $\varepsilon_3=\varepsilon_3^{\rm rms}$ in order to
make predictions for $v_3$.
It is however easier to
use the proportionality between $v_3$ and $\varepsilon_3$: one can
then do the calculation for an arbitrary $\varepsilon'_3$, and rescale
the final results by $\varepsilon_3^{\rm rms}/\varepsilon_3$.
We use $\varepsilon_3^{\rm rms}$ from the Monte-Carlo Glauber model
with Glauber initial conditions, and from the Monte-Carlo KLN model
with CGC initial conditions.

There is some arbitrariness in the definition of the triangularity
$\varepsilon_3$: one could for instance replace $r^2$ by $r^3$ in 
Eq.~(\ref{defepsilon3})~\cite{Li:2010}. 
With this replacement, both $\varepsilon_3$ and $\varepsilon_3^{\rm
  rms}$ (from the Monte-Carlo calculations) increase, but 
the ratio $\varepsilon_3^{\rm rms}/\varepsilon_3$ --- and therefore 
also our predicted $v_3$ --- changes little 
(less than $7\%$ for all centralities and both sets of initial 
conditions)\footnote{If one replaces $r^2$ by $r^k$ in
  Eq.~(\ref{defepsilon3}), $\varepsilon_n$ scales with $k$ like $k+2$
  for a smooth, symmetric density profile $\epsilon(r)$ deformed
  according to Eq.~(\ref{deformation}). Therefore, $\varepsilon_3$ is 
  larger by $\frac{5}{4}$ if defined with a factor $r^3$ instead of $r^2$.}.

Finally, we compute $v_3$ in viscous hydrodynamics.
It has been shown that RHIC data are fit equally well with
Glauber initial conditions and $\eta/s=0.08$ or with CGC initial
conditions and $\eta/s=0.16$~\cite{Luzum:2008cw}. 
The larger eccentricity of CGC (which should produce more elliptic flow)  is
compensated by the larger viscosity (larger damping and less flow), so
that the final values of $v_2$ are very similar.
For LHC energies, details are as in Ref.~\cite{Luzum:2009sb} (with
$v_3$ calculated from a Cooper-Frye freeze-out prescription).
In all cases, $v_3$ is found to be independent of the orientation of
the triangle $\psi'_3$.
In the case of Glauber initial conditions, we perform calculations of
$v_3$ with and without resonance decays at freeze
out~\cite{Sollfrank:1991xm}. Resonance decays roughly amount to multiplying
$v_3$ by $0.75$ at  RHIC, and by $0.83$  at LHC. Our CGC results are
computed without resonance decays, and multiplied by the same factor
at the end of the calculation.

\subsection{Results and comparison with data}
\begin{figure}[ht]
\includegraphics[width=\linewidth]{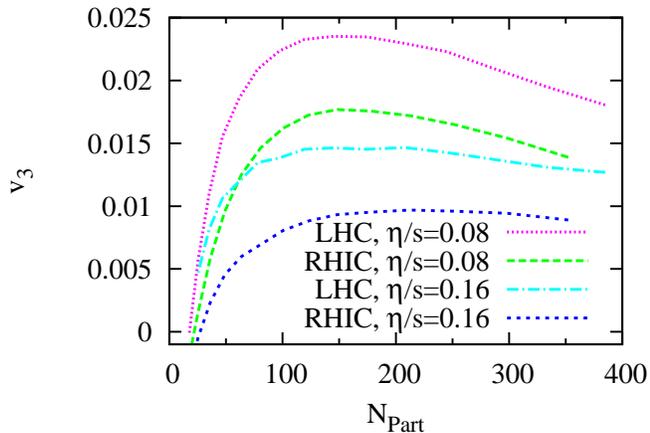}
\caption{(Color online) Average $v_3$ of pions as a function of the
  number of participants for Au-Au collisions at 200~GeV per nucleon
  (RHIC) and Pb-Pb collisions at 5.5~TeV per nucleon (LHC).  Hydrodynamic
  predictions are for Glauber initial conditions with $\eta/s$ = 0.08, and CGC
  initial conditions with $\eta/s$ = 0.16, which best fit $v_2$ data at RHIC~\cite{Luzum:2008cw}.}
\label{fig:v3predictions}
\end{figure}
Results are displayed in Fig.~\ref{fig:v3predictions} for both
sets of initial conditions. CGC initial conditions have both a smaller
triangularity, and a larger viscosity, so that they predict a much
smaller $v_3$. The change in viscosity explains roughly 70\%  of the
difference between CGC and Glauber at RHIC, and about half at LHC.
The centrality dependence is much flatter in
Fig.~\ref{fig:v3predictions} than in Fig.~\ref{fig:epsilon}.
The decrease of $\varepsilon_3^{\rm rms}$ with increasing $N_{\rm
  Part}$ is compensated by the increase of the system size and lifetime, which
leads to a smaller effective Knudsen number $K$ or, equivalently, a
smaller viscous correction.
We predict values of $v_3$ significantly larger at LHC than at RHIC.
This is because viscous damping is less important due to
the larger lifetime of the fluid at LHC~\cite{Luzum:2009sb}.

\begin{figure}[ht]
\includegraphics[width=\linewidth]{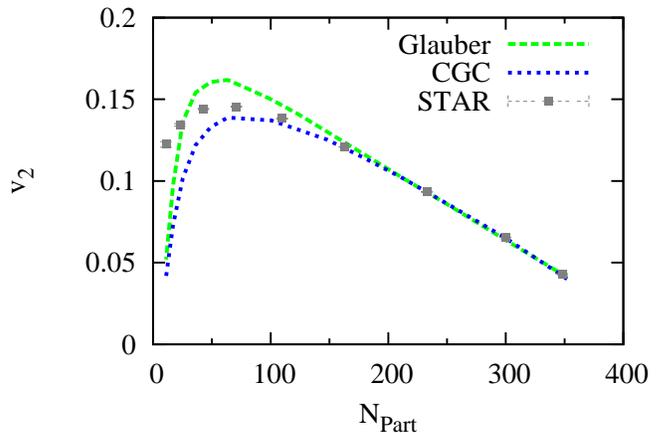}
\caption{(Color online)
$v_2$ for charged particles with $0.8<p_t<4$~GeV/c extracted from STAR
charge-independent correlation data~\cite{Abelev:2008un}, and
predictions from viscous hydrodynamics~\cite{Luzum:2008cw}  with
Glauber initial conditions and $\eta/s=0.08$, or CGC initial
conditions with $\eta/s=0.16$.
Theoretical calculations are for pions with the same $p_t$ cut as data,
and scaled by the rms eccentricity from the corresponding Monte-Carlo
model.  See text for details.}
\label{fig:starv2}
\end{figure}

\begin{figure}[ht]
\includegraphics[width=\linewidth]{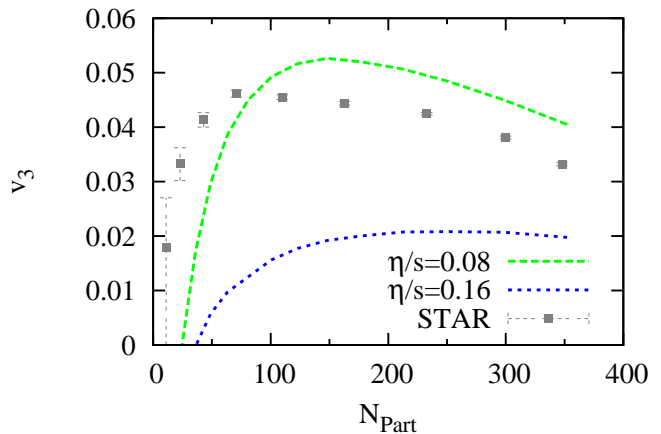}
\caption{(Color online)
Same as Fig.~\ref{fig:starv2} for $v_3$. }
\label{fig:starv3}
\end{figure}

Although experimental data for triangular flow are not yet available,
both $v_2$ and $v_3$ can be extracted from the measured two-particle
azimuthal correlation using
Eq.~(\ref{correlation})~\cite{Alver:2010gr}.
Figs.~\ref{fig:starv2} and \ref{fig:starv3} display a comparison
between  experimental data from STAR~\cite{Abelev:2008un} and our
hydrodynamic calculations.
The STAR data is obtained from correlations between particles at
midrapidity ($|\eta|<1$) and intermediate transverse momentum
($0.8<p_t<4.0$~GeV). The correlation results have been projected at
$1.2<\Delta\eta<1.9$ to reduce sensitivity to
nonflow effects.

We first discuss our results for $v_2$.
As explained above, our hydrodynamic model has smooth initial
conditions, and does not include the effect of eccentricity
fluctuations for $v_2$. Since $v_2\propto \varepsilon_2$ to a good
approximation, we have rescaled our result for $v_2$ by the rms
$\varepsilon_2$ from Fig.~\ref{fig:epsilon} (again using the Monte-Carlo Glauber
for the Glauber initial conditions and the Monte-Carlo KLN for CGC).
This rescaling significantly improves the agreement with data,
compared to~\cite{Luzum:2008cw}, for the most central bin.
As shown in Fig.~\ref{fig:starv2}, the agreement between theory and
data is excellent with both sets of initial conditions.
The smaller viscosity associated with Glauber initial conditions
results in a somewhat steeper centrality dependence than for CGC initial
conditions.

Results for $v_3$ are shown in Fig.~\ref{fig:starv3}.
The larger magnitude, compared to Fig.~\ref{fig:v3predictions}, is due
to the low $p_t$ cutoff. The cutoff also enhances the effect of
viscosity, resulting in a larger difference between Glauber and CGC.
With a low $p_t$ cutoff, the viscous correction is mostly due to
the distortion of the momentum distribution at
freeze-out~\cite{Teaney:2003kp}. The momentum dependence of this
distortion is strongly model-dependent~\cite{Dusling:2009df}.
The present calculation uses the standard quadratic ansatz,
which may overestimate the viscous correction at large
$p_t$~\cite{Luzum:2010ad}.
The magnitude and the centrality dependence of $v_3$ observed by STAR
are rather well reproduced by our calculation with Glauber initial
conditions, except for peripheral collisions where hydrodynamics is
not expected to be valid.

\begin{figure}[ht]
\includegraphics[width=\linewidth]{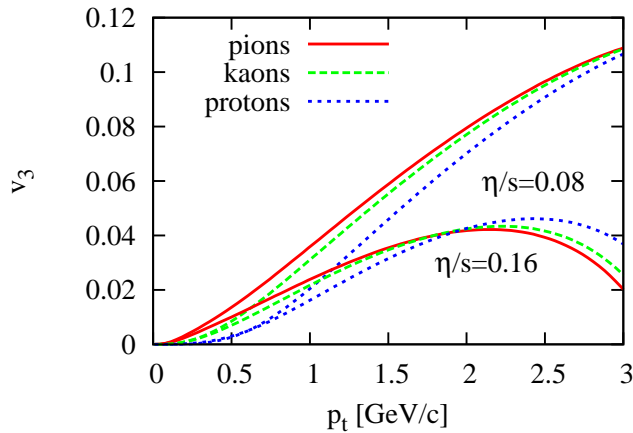}
\caption{(Color online)
Differential triangular flow for identified particles in central
($0-5\%$) Au-Au collisions at RHIC.}
\label{fig:identifiedv3}
\end{figure}

Fig.~\ref{fig:identifiedv3} displays our predictions for $v_3(p_t)$ of
identified particles at RHIC. As anticipated in
Ref.~\cite{Mishra:2008dm}, the well-known mass ordering of elliptic
flow \cite{Huovinen:2001cy} is also expected for $v_3$.
At high $p_t$, a strong viscous suppression is observed. As explained
above, the $p_t$ dependence of the viscous correction is model
dependent, and it is likely that the quadratic ansatz used here
overestimates the viscous corrections at large $p_t$~\cite{Luzum:2010ad}.
Note that effects of resonance decays are not included in
Fig.~\ref{fig:identifiedv3}. Resonance decays only change the results
slightly in the low-$p_t$ region.

\begin{figure}[ht]
\includegraphics[width=\linewidth]{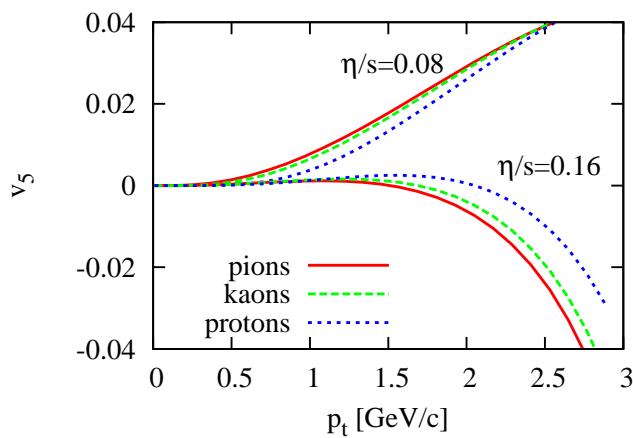}
\caption{(Color online)
Differential pentagonal flow for identified particles in central
($0-5\%$) Au-Au collisions at RHIC.}
\label{fig:identifiedv5}
\end{figure}

Finally, we have also computed $v_5$ along the same lines as
$v_3$. 
The driving force for $v_5$ is the rms $\varepsilon_5$, which
is very close to $\varepsilon_3$ (see
Fig.~\ref{fig:epsilon}). 
However, the hydrodynamic response is much
smaller, and viscous damping is also much larger as discussed in
Sec.~\ref{s:gaussian}. 
We find that the average integrated $v_5$ is smaller than
$v_3$ by at least a factor of 10. 
Results for differential $v_5$ are presented in
Fig.~\ref{fig:identifiedv5}. 
$v_5$ varies more strongly with $p_t$ than $v_2$ and $v_3$, and
becomes as large as $0.02$ at $p_t=1.5$~GeV/c if the $\eta/s$ is as
small as $0.08$. 
For larger viscosity, however, $v_5$ may be too small to measure 
even with a high $p_t$ trigger.

\section{Conclusions}
\label{s:conclusions}
We have presented a systematic study of triangular flow in ideal and
viscous hydrodynamics, and transport theory.
Triangular flow is driven by the average event-by-event triangularity in the
transverse distribution of nucleons, in the same way as
elliptic flow is driven by the initial eccentricity of this
distribution.
The physics of $v_3$ is in many respects similar to the physics of
$v_2$. In ideal hydrodynamics, the response to the initial deformation
is almost identical in both harmonics:
$v_3/\varepsilon_3\simeq v_2/\varepsilon_2\simeq 0.2$.
For quadrangular flow, $v_4/\varepsilon_4$ is smaller, typically by a
factor 2. 
For pentagonal flow, $v_5/\varepsilon_5$ is so small that $v_5$ is
unlikely to be measurable, even though $\varepsilon_3$ and
$\varepsilon_5$ are almost equal. 
$v_3$ develops slightly more slowly than $v_2$, though over comparable time scales.
The dependence on transverse momentum $p_t$ is similar
for $v_3$ and $v_2$, but $v_3/v_2$ increases with $p_t$.
Hydrodynamics predicts a similar mass ordering for $v_3(p_t)$ and
$v_2(p_t)$: $v_3$ at fixed $p_t$ is smaller for more massive
particles. These results can be checked experimentally by a differential measurement of triangular flow.

We have also made predictions for triangular flow, $v_3$, at RHIC and
LHC, using viscous hydrodynamics.
Using as input the triangularity from a standard Monte-Carlo Glauber
model, and a viscosity $\eta/s=0.08$, we reproduce both the
magnitude (within 20\%) and the centrality dependence of $v_3$
extracted from STAR correlation measurements, without any adjustable
parameter. Our results support the hypothesis made in
Ref.~\cite{Alver:2010gr} that triangular flow explains most of the
ridge and shoulder structures observed in the two-particle azimuthal
correlation.

Triangular flow is a sensitive probe of viscosity.
Viscous effects drive the energy and centrality dependence of $v_3$.
More central collisions have less fluctuations, hence smaller
triangularity. This decrease is to a large extent compensated by the
increase in the system size and lifetime, resulting in a very
slow decrease of $v_3$ with centrality
(except for peripheral collisions where viscous hydrodynamics is
unlikely to be valid).
Comparison with existing data favors a low value of $\eta/s$.
At LHC, smaller viscous corrections are expected due to the
increased lifetime of the fluid: we predict that $v_3$
should be  larger than at RHIC, typically by a factor $\frac{4}{3}$.

The absolute value of $v_3$ scales linearly with the average initial
initial triangularity. We have used two models of initial geometry which incorporate
fluctuations, the Monte-Carlo Glauber model and the Monte-Carlo KLN
model. The underlying source of fluctuations is the
same in both of these models.
More work is needed to constrain initial fluctuations on the
theoretical side.
More work is also needed to incorporate these fluctuations more readily into
hydrodynamic calculations.
Although triangular flow is expected to be created by lumpy initial
conditions, our predictions are based on smooth initial
conditions, in the same spirit as the study of transverse momentum
fluctuations of Ref.~\cite{Broniowski:2009fm}.
The underlying assumption is that
$v_3/\varepsilon_3$ is the same for lumpy initial conditions and for
smooth initial conditions. The validity of this assumption should
eventually be checked. 

Triangular flow is a new observable which should be used to constrain
models of heavy-ion collisions, along with elliptic flow.  
Elliptic flow depends on initial eccentricity, fluctuations,
and viscosity, which are poorly constrained theoretically.
Triangular flow solely depends on fluctuations and
viscosity, with a stronger sensitivity to viscosity than $v_2$.
Two different sets of initial conditions, which fit $v_2$ data equally
well, give very different results for $v_3$.
Experiments could measure $v_3$ as a function of transverse
momentum, system size and centrality.
As shown in this paper, theoretical predictions for the dependence of
$v_3$ on these parameters are very specific.
If experiments confirm our predictions, simultaneous analyses of $v_2$
and $v_3$ can be used to improve our understanding of the initial
geometry of heavy-ion collisions, and pin down the viscosity of hot QCD.

\section*{Acknowledgments}

This work is funded by ``Agence Nationale de la Recherche'' under grant
ANR-08-BLAN-0093-01 and by U.S. DOE grant DE-FG02-94ER40818.
We thank S. Gavin, T. Hirano, P. Huovinen and A. Poskanzer for
stimulating discussions, and W. Zajc for useful comments on the manuscript.
M. L. and J.-Y. O. thank the organizers of the program ``Quantifying
the properties of hot QCD matter'' and the Institute for Nuclear Theory at the
University of Washington, where part of this work was done, for its
hospitality, and the Department of Energy for partial support.


\begin{thebibliography}{99}


\bibitem{Ackermann:2000tr}
  K.~H.~Ackermann {\it et al.}  [STAR Collaboration],
  Phys.\ Rev.\ Lett.\  {\bf 86}, 402 (2001)
  [arXiv:nucl-ex/0009011].


\bibitem{Adare:2008cqb}
  A.~Adare {\it et al.}  [PHENIX Collaboration],
  Phys.\ Rev.\  C {\bf 78}, 014901 (2008)
  [arXiv:0801.4545 [nucl-ex]].

\bibitem{Abelev:2008nda}
  B.~I.~Abelev {\it et al.}  [STAR Collaboration],
  Phys.\ Rev.\ Lett.\  {\bf 102}, 052302 (2009)
  [arXiv:0805.0622 [nucl-ex]].

\bibitem{Abelev:2008un}
  B.~I.~Abelev {\it et al.}  [STAR Collaboration],
  arXiv:0806.0513 [nucl-ex].

\bibitem{Alver:2008gk}
  B.~Alver {\it et al.}  [PHOBOS Collaboration],
  Phys.\ Rev.\  C {\bf 81}, 024904 (2010)
  [arXiv:0812.1172 [nucl-ex]].

\bibitem{Alver:2009id}
  B.~Alver {\it et al.}  [PHOBOS Collaboration],
  Phys.\ Rev.\ Lett.\  {\bf 104}, 062301 (2010)
  [arXiv:0903.2811 [nucl-ex]].


\bibitem{Abelev:2009qa}
  B.~I.~Abelev {\it et al.}  [STAR Collaboration],
  Phys.\ Rev.\  C {\bf 80}, 064912 (2009)
  [arXiv:0909.0191 [nucl-ex]].


\bibitem{Dumitru:2008wn}
  A.~Dumitru, F.~Gelis, L.~McLerran and R.~Venugopalan,
  Nucl.\ Phys.\  A {\bf 810}, 91 (2008)
  [arXiv:0804.3858 [hep-ph]].


\bibitem{Alver:2010gr}
  B.~Alver and G.~Roland,
  Phys.\ Rev.\  C {\bf 81}, 054905 (2010)
  [arXiv:1003.0194 [nucl-th]].

\bibitem{Luzum:2008cw}
  M.~Luzum and P.~Romatschke,
  Phys.\ Rev.\  C {\bf 78}, 034915 (2008)
  [Erratum-ibid.\  C {\bf 79}, 039903 (2009)]
  [arXiv:0804.4015 [nucl-th]].

\bibitem{Gombeaud:2007ub}
  C.~Gombeaud and J.~Y.~Ollitrault,
  Phys.\ Rev.\  C {\bf 77}, 054904 (2008)
  [arXiv:nucl-th/0702075].

\bibitem{Borghini:2005kd}
  N.~Borghini and J.~Y.~Ollitrault,
  Phys.\ Lett.\  B {\bf 642}, 227 (2006)
  [arXiv:nucl-th/0506045].

\bibitem{Luzum:2011}
M.~Luzum, C.~Gombeaud and J.-Y.~Ollitrault, in preparation.

\bibitem{Voloshin:1994mz}
  S.~Voloshin and Y.~Zhang,
  Z.\ Phys.\  C {\bf 70}, 665 (1996)
  [arXiv:hep-ph/9407282].

\bibitem{Poskanzer:1998yz}
  A.~M.~Poskanzer and S.~A.~Voloshin,
  Phys.\ Rev.\  C {\bf 58}, 1671 (1998)
  [arXiv:nucl-ex/9805001].

\bibitem{Wang:1991qh}
  S.~Wang {\it et al.},
  Phys.\ Rev.\  C {\bf 44}, 1091 (1991).

\bibitem{Ollitrault:2009ie}
  J.~Y.~Ollitrault, A.~M.~Poskanzer and S.~A.~Voloshin,
  Phys.\ Rev.\  C {\bf 80}, 014904 (2009)
  [arXiv:0904.2315 [nucl-ex]].

\bibitem{Alver:2008zza}
  B.~Alver {\it et al.},
  Phys.\ Rev.\  C {\bf 77}, 014906 (2008)
  [arXiv:0711.3724 [nucl-ex]].


\bibitem{Huovinen:2007xh}
  P.~Huovinen,
  Eur.\ Phys.\ J.\  A {\bf 37}, 121 (2008)
  [arXiv:0710.4379 [nucl-th]].

\bibitem{Hirano:2009ah}
  T.~Hirano and Y.~Nara,
  Phys.\ Rev.\  C {\bf 79}, 064904 (2009)
  [arXiv:0904.4080 [nucl-th]].

\bibitem{Song:2009rh}
  H.~Song and U.~W.~Heinz,
  Phys.\ Rev.\  C {\bf 81}, 024905 (2010)
  [arXiv:0909.1549 [nucl-th]].

\bibitem{Broniowski:2009fm}
  W.~Broniowski, M.~Chojnacki and L.~Obara,
  Phys.\ Rev.\  C {\bf 80}, 051902 (2009)
  [arXiv:0907.3216 [nucl-th]].

\bibitem{Bozek:2009dw}
  P.~Bozek,
  Phys.\ Rev.\  C {\bf 81}, 034909 (2010)
  [arXiv:0911.2397].

\bibitem{Miller:2003kd}
  M.~Miller and R.~Snellings,
  arXiv:nucl-ex/0312008.

\bibitem{Manly:2005zy}
  S.~Manly {\it et al.}  [PHOBOS Collaboration],
  Nucl.\ Phys.\  A {\bf 774}, 523 (2006)
  [arXiv:nucl-ex/0510031].

\bibitem{Gyulassy:1996br}
  M.~Gyulassy, D.~H.~Rischke and B.~Zhang,
  Nucl.\ Phys.\ A {\bf 613}, 397 (1997)
  [arXiv:nucl-th/9609030].


\bibitem{Socolowski:2004hw}
  O.~J.~Socolowski, F.~Grassi, Y.~Hama and T.~Kodama,
  Phys.\ Rev.\ Lett.\  {\bf 93}, 182301 (2004)
  [arXiv:hep-ph/0405181].

\bibitem{Holopainen:2010gz}
  H.~Holopainen, H.~Niemi and K.~J.~Eskola,
  arXiv:1007.0368.

\bibitem{Werner:2010aa}
  K.~Werner, I.~Karpenko, T.~Pierog, M.~Bleicher and K.~Mikhailov,
  arXiv:1004.0805.


\bibitem{Mishra:2007tw}
  A.~P.~Mishra, R.~K.~Mohapatra, P.~S.~Saumia and A.~M.~Srivastava,
  Phys.\ Rev.\  C {\bf 77}, 064902 (2008)
  [arXiv:0711.1323 [hep-ph]].

\bibitem{Andrade:2006yh}
  R.~Andrade, F.~Grassi, Y.~Hama, T.~Kodama and O.~J.~Socolowski,
  Phys.\ Rev.\ Lett.\  {\bf 97}, 202302 (2006)
  [arXiv:nucl-th/0608067].

\bibitem{Alver:2006wh}
  B.~Alver {\it et al.}  [PHOBOS Collaboration],
  Phys.\ Rev.\ Lett.\  {\bf 98}, 242302 (2007)
  [arXiv:nucl-ex/0610037].

\bibitem{Gombeaud:2009ye}
  C.~Gombeaud and J.~Y.~Ollitrault,
  Phys.\ Rev.\  C {\bf 81}, 014901 (2010)
  [arXiv:0907.4664 [nucl-th]].

\bibitem{Alver:2010rt}
  B.~Alver {\it et al.}  [PHOBOS Collaboration],
  Phys.\ Rev.\  C {\bf 81}, 034915 (2010)
  [arXiv:1002.0534 [nucl-ex]].


\bibitem{Ollitrault:1992bk}
  J.~Y.~Ollitrault,
  Phys.\ Rev.\  D {\bf 46}, 229 (1992).

\bibitem{Sollfrank:1991xm}
  J.~Sollfrank, P.~Koch and U.~W.~Heinz,
  Z.\ Phys.\  C {\bf 52}, 593 (1991).

\bibitem{Bhalerao:2005mm}
  R.~S.~Bhalerao, J.~P.~Blaizot, N.~Borghini and J.~Y.~Ollitrault,
  Phys.\ Lett.\  B {\bf 627}, 49 (2005)
  [arXiv:nucl-th/0508009].

\bibitem{Sorge:1998mk}
  H.~Sorge,
  Phys.\ Rev.\ Lett.\  {\bf 82}, 2048 (1999)
  [arXiv:nucl-th/9812057].

\bibitem{otherhydro} P. Huovinen, T. Hirano, private communications.

\bibitem{Dinh:1999mn}
  P.~M.~Dinh, N.~Borghini and J.~Y.~Ollitrault,
  Phys.\ Lett.\  B {\bf 477}, 51 (2000)
  [arXiv:nucl-th/9912013].

\bibitem{Mishra:2008dm}
  A.~P.~Mishra, R.~K.~Mohapatra, P.~S.~Saumia and A.~M.~Srivastava,
  Phys.\ Rev.\  C {\bf 81}, 034903 (2010)
  [arXiv:0811.0292 [hep-ph]].

\bibitem{Putschke:2007mi}
  J.~Putschke,
  J.\ Phys.\ G {\bf 34}, S679 (2007)
  [arXiv:nucl-ex/0701074].

\bibitem{Daugherity:2008su}
  M.~Daugherity  [STAR Collaboration],
  J.\ Phys.\ G {\bf 35}, 104090 (2008)
  [arXiv:0806.2121 [nucl-ex]].

\bibitem{Drescher:2007cd}
  H.~J.~Drescher, A.~Dumitru, C.~Gombeaud and J.~Y.~Ollitrault,
  Phys.\ Rev.\  C {\bf 76}, 024905 (2007)
  [arXiv:0704.3553 [nucl-th]].

\bibitem{Voloshin:1999gs}
  S.~A.~Voloshin and A.~M.~Poskanzer,
  Phys.\ Lett.\  B {\bf 474}, 27 (2000)
  [arXiv:nucl-th/9906075].

\bibitem{Gavin:2010}
Sean Gavin, private communication.

\bibitem{Nagle:2009ip}
  J.~L.~Nagle, P.~Steinberg and W.~A.~Zajc,
  Phys.\ Rev.\  C {\bf 81}, 024901 (2010)
  [arXiv:0908.3684 [nucl-th]].

\bibitem{Heinz:2009cv}
  U.~W.~Heinz, J.~S.~Moreland and H.~Song,
  Phys.\ Rev.\  C {\bf 80}, 061901 (2009)
  [arXiv:0908.2617 [nucl-th]].

\bibitem{Kovtun:2004de}
  P.~Kovtun, D.~T.~Son and A.~O.~Starinets,
  Phys.\ Rev.\ Lett.\  {\bf 94}, 111601 (2005)
  [arXiv:hep-th/0405231].

\bibitem{Alver:2008aq}
  B.~Alver, M.~Baker, C.~Loizides and P.~Steinberg,
  arXiv:0805.4411 [nucl-ex].

\bibitem{Bozek:2010wt}
  P.~Bozek, M.~Chojnacki, W.~Florkowski and B.~Tomasik,
  arXiv:1007.2294 [Unknown].

\bibitem{Drescher:2007ax}
  H.~J.~Drescher and Y.~Nara,
  Phys.\ Rev.\  C {\bf 76}, 041903 (2007)
  [arXiv:0707.0249 [nucl-th]].

\bibitem{Miller:2007ri}
  M.~L.~Miller, K.~Reygers, S.~J.~Sanders and P.~Steinberg,
  Ann.\ Rev.\ Nucl.\ Part.\ Sci.\  {\bf 57}, 205 (2007)
  [arXiv:nucl-ex/0701025].

\bibitem{Lappi:2006xc}
  T.~Lappi and R.~Venugopalan,
  Phys.\ Rev.\  C {\bf 74}, 054905 (2006)
  [arXiv:nucl-th/0609021].

\bibitem{Bhalerao:2006tp}
  R.~S.~Bhalerao and J.~Y.~Ollitrault,
  Phys.\ Lett.\  B {\bf 641}, 260 (2006)
  [arXiv:nucl-th/0607009].

\bibitem{Li:2010}
Yan Li and D.Teaney, ``Triangle and Dipole Flow in Ideal 
Hydrodynamics'', in preparation; D. Teaney, contribution to
Strong and Electroweak Matter 2010,
June 29 - July 2, 2010, Montreal, Canada.

\bibitem{Luzum:2009sb}
  M.~Luzum and P.~Romatschke,
  Phys.\ Rev.\ Lett.\  {\bf 103}, 262302 (2009)
  [arXiv:0901.4588 [nucl-th]].

\bibitem{Teaney:2003kp}
  D.~Teaney,
  Phys.\ Rev.\  C {\bf 68}, 034913 (2003)
  [arXiv:nucl-th/0301099].

\bibitem{Dusling:2009df}
  K.~Dusling, G.~D.~Moore and D.~Teaney,
  Phys.\ Rev.\  C {\bf 81}, 034907 (2010)
  [arXiv:0909.0754 [nucl-th]].

\bibitem{Luzum:2010ad}
  M.~Luzum and J.~Y.~Ollitrault,
  Phys.\ Rev.\  C {\bf 82}, 014906 (2010)
  [arXiv:1004.2023 [nucl-th]].

\bibitem{Huovinen:2001cy}
  P.~Huovinen, P.~F.~Kolb, U.~W.~Heinz, P.~V.~Ruuskanen and S.~A.~Voloshin,
  Phys.\ Lett.\  B {\bf 503}, 58 (2001)
  [arXiv:hep-ph/0101136].
 
\end{thebibliography}
\end{document}